# Macroscopic entanglement of three magnon modes in three cavities via optical parametric amplifier

YING ZHOU,[1,*] AND GUOQIANG ZHANG[2]

[1] *School of Materials Science and Engineering, Taizhou University, Taizhou 318000, China*
[2] *School of Physics, Hangzhou Normal University, Hangzhou 311121, China*
*\*zying@tzc.edu.cn*

**Abstract:** We propose a scheme to generate bipartite and tripartite entanglements of three magnon modes in a three-cavity system using a nonlinear optical parametric amplifier (OPA). The three magnon modes in three YIG spheres are respectively placed inside three cavities near the maximum magnetic fields of the cavities and coupled to cavity modes via linear magnetic dipole interaction. Additionally, linear coupling interaction exists between two cavities. Using experimentally feasible parameters, we demonstrate that OPA can prepare the three magnon modes in a steady-state entangled state, bipartite and tripartite entanglements increase with the nonlinear interaction strength of OPA. An alternative approach to enhance quantum entanglement involves multiplexed OPA inputs. By employing individual OPA for each cavity, we observe a significant improvement in entanglement generation. All the entanglements are robust against bath temperature.

## 1. Introduction

Recent years have witnessed significant progress in studying the light-matter interactions in ferrimagnetic systems[1], especially in the systems including yttrium iron garnet (YIG) material, owing to its low dissipation rate and high spin density. Studies have shown that the Kittel mode[2] in YIG sphere can achieve strong coupling[3-8] even ultra-strong coupling[9, 10] with microwave photons in cavity-magnon system, giving rise to cavity-magnon polaritions. Numerous intriguing phenomena have been explored in these systems, including quantum entanglement[11-24], squeezed states[25-28], bistability[29], manipulation of distant spin currents[30], magnon gradient memory[31], quantum phase transition[32] and the exceptional point[33, 34]. Among them, the entanglement between magnons, photons, and phonons has attracted particular attention due to its broad applications in quantum information science[35-38] and quantum networks[39].

To generate entanglement, the system must possess two key elements: coupling interaction and nonlinearity. The predominant nonlinearity employed in cavity-magnon systems is magnetostrictive interaction, which has enabled the generation of various entangled states, including magnon-photon-phonon entanglement[11], magnon-magnon entanglement[14], photon-photon entanglement[15], phonon-phonon entanglement[13], phonon-magnon entanglement[16] and microwave-optics entanglement[12], etc. Another nonlinearity used is Kerr nonlinearity[17], and magnon-magnon entangled state can be generated. The other two nonlinearities are Josephson parametric amplifier (JPA)[18] and optical parametric amplifier (OPA)[19, 20], magnon-magnon entanglement and photon-photon entanglement are also found. Furthermore, studies have shown that combining two or three different nonlinearities[21-24], the cavity-magnon QED system can significantly enhance entanglement generation. Notably, all of these entanglement studies are based on one or two cavities, and little research have been conducted on the entanglement of a three-cavity or even multi-cavity system.

In this paper, we consider preparing macroscopic entangled state of three YIG spheres inside a three-cavity QED system by exploiting nonlinear OPA. The Kittel mode of YIG sphere couples to the cavity mode via linear beam-splitter interaction, and the coupling interaction between two cavities is also linear. Notably, apart from the OPA nonlinearity, no additional nonlinear factors are introduced in our cavity-magnon QED system. We find that the three magnon modes are entangled in the steady state, and both the bipartite and tripartite entanglements increase with multiplexed OPA inputs.

## 2. System model

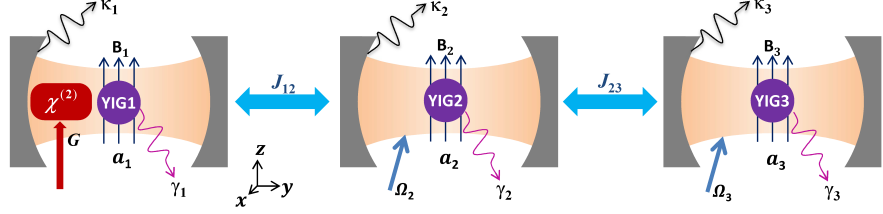

**Fig. 1.** Sketch of the cavity-magnon QED system. A cavity-magnon QED system consists of three microwave cavities with coupling strengths $J_{12}$ and $J_{23}$. Three YIG spheres are respectively placed inside the cavities near the maximum magnetic fields of the cavities and simultaneously subjected to uniform bias magnetic fields, resulting in the resonance frequencies $\omega_{m_1}$, $\omega_{m_2}$ and $\omega_{m_3}$. The magnetic fields of the cavity modes and the bias magnetic fields are along the $x-$axis and $z-$axis, respectively. The cavity 1 is driven by an optical parametric amplifier (OPA) with nonlinear interaction strength $G$, the cavity 2 and cavity 3 are respectively driven by microwave fields with Rabi frequency $\Omega_j$ ($j=2,3$). The decay rates of cavity modes($a_1$, $a_2$, $a_3$) and magnon modes ($m_1$, $m_2$, $m_3$) are given by $\kappa_1$, $\kappa_2$, $\kappa_3$, $\gamma_1$, $\gamma_2$ and $\gamma_3$, respectively.

The system consists of three microwave cavities and three magnon modes in three YIG spheres. Each YIG sphere is placed inside the cavity near the maximum magnetic field of the cavity and simultaneously subjected to a uniform bias magnetic field, as shown in Fig. 1. The magnon mode(e.g., the Kittel mode[2]) is embodied by the collective motion of a large number of spins in the YIG crystal, and couples to the microwave cavity photons via magnetic dipole interaction. Owing to the high spin density of YIG material, this coupling can reach the strong-coupling regime for $g_j \propto \sqrt{N_{m_j}}$ [5] with $N_{m_j}$ the total number of spins in the $j$-th YIG sphere. Since the size of YIG sphere used in our system is much smaller than the microwave wavelength, radiation pressure effects on the YIG sphere can be safely neglected. The Hamiltonian of such a cavity-magnon QED system reads

$$H/\hbar = \sum_{j=1,2,3}\omega_{a_j}a_j^\dagger a_j + \sum_{j=1,2,3}\omega_{m_j}m_j^\dagger m_j + g_1(a_1+a_1^\dagger)(m_1+m_1^\dagger) + g_2(a_2+a_2^\dagger)(m_2+m_2^\dagger)$$
$$+ g_3(a_3+a_3^\dagger)(m_3+m_3^\dagger) + J_{12}(a_1^\dagger a_2 + a_1 a_2^\dagger) + J_{23}(a_2^\dagger a_3 + a_2 a_3^\dagger) \quad (1)$$
$$+ iG(a_1^{\dagger 2}e^{-2i\omega_d t} - a_1^2 e^{2i\omega_d t}) + i\Omega_2(a_2^\dagger e^{-i\omega_d t} - a_2 e^{i\omega_d t}) + i\Omega_3(a_3^\dagger e^{-i\omega_d t} - a_3 e^{i\omega_d t})$$

where $a_j$ and $a_j^\dagger$ ($j=1,2,3$) denote the annihilation and creation operators of cavity mode $j$ with resonance frequencies $\omega_{a_j}$. Similarly, $m_j$ and $m_j^\dagger$ are the annihilation and creation operators of spatially uniform Kittel mode $j$ with resonance frequencies $\omega_{m_j}$, which is determined by the external bias magnetic field (along the $z$ directions) via $\omega_{m_j} = \gamma H_j$ and can be flexibly adjusted, where $\gamma/2\pi = 28$GHz/T is the gyromagnetic ratio. We have $[O,O^\dagger]=1$ ($O=a_j,m_j$). $g_j$ denotes the linear coupling rate between cavity mode $a_j$ and

Kittel mode $m_j$, which can be much larger than the decay rates of cavity mode $\kappa_j$ and magnon mode $\gamma_j$, i.e. $g_j > \kappa_j, \gamma_j$ ($j=1,2,3$). Therefore, the cavity-magnon QED system is in the strong coupling regime [3-8] and the rotation-wave approximation can be applied. Meanwhile, $J_{12}$ ($J_{23}$) is the linear coupling strength of two microwave cavities. To enable entanglement generation in the system, a nonlinear driving is adopted by optical parametric amplifier (OPA) with nonlinear interaction strength $G$ in cavity 1. Additionally, cavity 2 and cavity 3 are respectively driven by linear microwave fields with the Rabi frequency $\Omega_j$ ($j=2,3$), $\Omega_j = \sqrt{2P_d \kappa_j / \hbar \omega_{a_j}}$, here, $P_d$ is the power of drive field and $\kappa_j$ represents the decay rate of cavity field $j$ with frequency $\omega_{a_j}$.

Under the rotation-wave approximation, the linear coupling term of cavity mode and magnon mode $g_j(a_j + a_j^\dagger)(m_j + m_j^\dagger)$ becomes $g_j(a_j^\dagger m_j + a_j m_j^\dagger)$ [3-7, 11, 40]. Due to this linear coupling, the squeezing from the cavity photon can be transferred to the magnon, which is referred to as the beam-splitter interaction[11]. In the frame rotating at the drive frequency $\omega_d$, the Hamiltonian of our system can be written as

$$H/\hbar = \Delta_{a_1} a_1^\dagger a_1 + \Delta_{a_2} a_2^\dagger a_2 + \Delta_{a_3} a_3^\dagger a_3 + \Delta_{m_1} m_1^\dagger m_1 + \Delta_{m_2} m_2^\dagger m_2 + \Delta_{m_3} m_3^\dagger m_3 + g_1\left(a_1^\dagger m_1 + a_1 m_1^\dagger\right)$$
$$+ g_2\left(a_2^\dagger m_2 + a_2 m_2^\dagger\right) + g_3\left(a_3^\dagger m_3 + a_3 m_3^\dagger\right) + J_{12}(a_1^\dagger a_2 + a_1 a_2^\dagger) + J_{23}(a_2^\dagger a_3 + a_2 a_3^\dagger) \quad (2)$$
$$+ iG\left(a_1^{\dagger 2} - a_1^2\right) + i\Omega_2\left(a_2^\dagger - a_2\right) + i\Omega_3\left(a_3^\dagger - a_3\right)$$

where $\Delta_{a_j} = \omega_{a_j} - \omega_d$ is the detuning of cavity mode $j$ ($j=1,2,3$), and $\Delta_{m_j} = \omega_{m_j} - \omega_0$ is the detuning of magnon mode $j$. Considering the dissipations and the input noises, the quantum Langevin equations (QLEs)[41] for the cavity-magnon QED system are given by

$$\dot{a}_1 = -\left(i\Delta_{a_1} + \kappa_1\right)a_1 - ig_1 m_1 - iJ_{12} a_2 + 2G a_1^\dagger + \sqrt{2\kappa_1}\, a_1^{in}$$
$$\dot{a}_2 = -\left(i\Delta_{a_2} + \kappa_2\right)a_2 - ig_2 m_2 - iJ_{12} a_1 - iJ_{23} a_3 + \Omega_2 + \sqrt{2\kappa_2}\, a_2^{in}$$
$$\dot{a}_3 = -\left(i\Delta_{a_3} + \kappa_3\right)a_3 - ig_3 m_3 - iJ_{23} a_2 + \Omega_3 + \sqrt{2\kappa_3}\, a_3^{in} \quad (3)$$
$$\dot{m}_1 = -\left(i\Delta_{m_1} + \gamma_1\right)m_1 - ig_1 a_1 + \sqrt{2\gamma_1}\, m_1^{in}$$
$$\dot{m}_2 = -\left(i\Delta_{m_2} + \gamma_2\right)m_2 - ig_2 a_2 + \sqrt{2\gamma_2}\, m_2^{in}$$
$$\dot{m}_3 = -\left(i\Delta_{m_3} + \gamma_3\right)m_3 - ig_3 a_3 + \sqrt{2\gamma_3}\, m_3^{in}$$

where $a_j^{in}$ ($m_j^{in}$) with zero mean value is the input noise operator for the $j$th cavity mode (magnon mode), characterized by quantum correlations as described by the correlation functions[42]

$$\langle a_j^{in}(t) a_j^{in\dagger}(t')\rangle = \left[N_{a_j}(\omega_{a_j}) + 1\right]\delta(t-t'),$$
$$\langle a_j^{in\dagger}(t) a_j^{in}(t')\rangle = N_{a_j}(\omega_{a_j})\delta(t-t');$$
$$\langle m_j^{in}(t) m_j^{in\dagger}(t')\rangle = \left[N_{m_j}(\omega_{m_j}) + 1\right]\delta(t-t'), \quad (4)$$
$$\langle m_j^{in\dagger}(t) m_j^{in}(t')\rangle = N_{m_j}(\omega_{m_j})\delta(t-t'); (j=1,2,3)$$

where $N_l(\omega_l) = \left[\exp(\hbar\omega_l / k_B T) - 1\right]^{-1}$ ($l = a_1, a_2, a_3, m_1, m_2, m_3$) are the equilibrium mean thermal photon numbers and magnon numbers, respectively. $k_B$ is the Boltzmann constant and $T$ denotes the environmental temperature.

Since cavity 1 is strongly driving by OPA, cavity 2 and cavity 3 are respectively strong driven by linear microwave fields, resulting in large amplitudes of cavity mode $|\langle a_j \rangle| \gg 1$ and magnon mode $|\langle m_j \rangle| \gg 1$ ($j=1,2,3$) in the steady state, which enables us to linearize the system dynamics and neglect small second-order fluctuation terms, the operators of cavity modes and magnon modes can be expressed as $O = \langle O \rangle + \delta O$ ($O = a_1, a_2, a_3, m_1, m_2, m_3$). Here, $\langle O \rangle$ is the mean value of operator $O$ and $\delta O$ denotes the zero-mean quantum fluctuation. Two sets of equations can be derived: one for semi-classical mean values and the other for quantum fluctuations. From the former set of equations, we can obtain mean photon numbers and mean magnon numbers in the steady state.

The later set of equations is the dynamic of quantum fluctuations of the system related to entanglement and squeezing. The quadrature fluctuations of photons and magnons can be defined as $\delta X_j = (\delta a_j + \delta a_j^\dagger)/\sqrt{2}$, $\delta Y_j = i(\delta a_j^\dagger - \delta a_j)/\sqrt{2}$, $\delta x_j = (\delta m_j + \delta m_j^\dagger)/\sqrt{2}$, and $\delta y_j = i(\delta m_j^\dagger - \delta m_j)/\sqrt{2}$ ($j=1,2,3$). Then the evolution of quadrature fluctuations can be described by linearized QLEs in matrix form

$$\dot{f}(t) = Af(t) + \eta \tag{5}$$

where $f(t) = [\delta X_1, \delta Y_1, \delta X_2, \delta Y_2, \delta X_3, \delta Y_3, \delta x_1, \delta y_1, \delta x_2, \delta y_2, \delta x_3, \delta y_3]^T$ denotes the vector of fluctuation operators, and $\eta(t) = [\sqrt{2\kappa_1}X_1^{in}, \sqrt{2\kappa_1}Y_1^{in}, \sqrt{2\kappa_2}X_2^{in}, \sqrt{2\kappa_2}Y_2^{in}, \sqrt{2\kappa_3}X_3^{in}, \sqrt{2\kappa_3}Y_3^{in},$ $\sqrt{2\gamma_1}x_1^{in}, \sqrt{2\gamma_1}y_1^{in}, \sqrt{2\gamma_2}x_2^{in}, \sqrt{2\gamma_2}y_2^{in}, \sqrt{2\gamma_3}x_3^{in}, \sqrt{2\gamma_3}y_3^{in}]^T$ is the vector of noise operators. The drift matrix A can be expressed as

$$A = \begin{pmatrix}
2G-\kappa_1 & \Delta_{a_1} & 0 & J_{12} & 0 & 0 & 0 & g_1 & 0 & 0 & 0 & 0 \\
-\Delta_{a_1} & -2G-\kappa_1 & -J_{12} & 0 & 0 & 0 & -g_1 & 0 & 0 & 0 & 0 & 0 \\
0 & J_{12} & -\kappa_2 & \Delta_{a_2} & 0 & J_{23} & 0 & 0 & 0 & g_2 & 0 & 0 \\
-J_{12} & 0 & -\Delta_{a_2} & -\kappa_2 & -J_{23} & 0 & 0 & 0 & -g_2 & 0 & 0 & 0 \\
0 & 0 & 0 & J_{23} & -\kappa_3 & \Delta_{a_3} & 0 & 0 & 0 & 0 & 0 & g_3 \\
0 & 0 & -J_{23} & 0 & -\Delta_{a_3} & -\kappa_3 & 0 & 0 & 0 & 0 & -g_3 & 0 \\
0 & g_1 & 0 & 0 & 0 & 0 & -\gamma_1 & \Delta_{m_1} & 0 & 0 & 0 & 0 \\
-g_1 & 0 & 0 & 0 & 0 & 0 & -\Delta_{m_1} & -\gamma_1 & 0 & 0 & 0 & 0 \\
0 & 0 & 0 & g_2 & 0 & 0 & 0 & 0 & -\gamma_2 & \Delta_{m_2} & 0 & 0 \\
0 & 0 & -g_2 & 0 & 0 & 0 & 0 & 0 & -\Delta_{m_2} & -\gamma_2 & 0 & 0 \\
0 & 0 & 0 & 0 & 0 & g_3 & 0 & 0 & 0 & 0 & -\gamma_3 & \Delta_{m_3} \\
0 & 0 & 0 & 0 & -g_3 & 0 & 0 & 0 & 0 & 0 & -\Delta_{m_3} & -\gamma_3
\end{pmatrix} \tag{6}$$

Due to the linearized system dynamics and the Gaussian nature of all quantum noises in our cavity-magnon QED system, the steady-state of quantum fluctuations form a continuous variable six-mode Gaussian state and can be completely characterized by a $12 \times 12$ covariance matrix $V$, here, $V_{ij} = \langle f_i(t)f_j(t') + f_j(t')f_i(t) \rangle/2$ ($i,j = 1,2,\ldots,12$). The stationary covariance matrix $V$ can be obtained directly by solving the Lyapunov equation[43, 44]

$$AV + VA^T = -D \tag{7}$$

where $D = \text{diag}[\kappa_1(2N_{a_1}+1), \kappa_1(2N_{a_1}+1), \kappa_2(2N_{a_2}+1), \kappa_2(2N_{a_2}+1), \kappa_3(2N_{a_3}+1), \kappa_3(2N_{a_3}+1), \gamma_1(2N_{m_1}+1), \gamma_1(2N_{m_1}+1), \gamma_2(2N_{m_2}+1), \gamma_2(2N_{m_2}+1), \gamma_3(2N_{m_3}+1), \gamma_3(2N_{m_3}+1)]$ is the diffusion matrix, which is defined as $D_{ij}\delta(t-t') = \langle \eta_i(t)\eta_j(t') + \eta_j(t')\eta_i(t)\rangle/2$。Using the covariance matrix $V$, we can calculate entanglement and squeezing. We adopt the logarithmic negativity $E_N$ [45, 46] to quantify the bipartite entanglement and the minimum residual contangle $R_\tau^{min}$ [47, 48] to quantify the tripartite entanglement.

The logarithmic negativity[45, 46] is given by
$$E_N \equiv \max[0, -\ln 2\tilde{v}_-] \tag{8}$$

Where $\hat{v}_- = \min[\text{eig}|i\Omega_2 \tilde{V}_4|]$ is the minimum symplectic eigenvalue of covariance matrix $\tilde{V}_4$, here, $\tilde{V}_4 = P_{1|2}V_4 P_{1|2}$ with $P_{1|2} = \text{diag}(1,-1,1,1)$, and $V_4$ is a $4\times 4$ covariance matrix of two arbitrary modes in our six-mode system, which can be obtained by directly removing rows and columns corresponding to irrelevant modes from the full covariance matrix $V$. $\Omega_2 = \oplus_{j=1}^2 i\sigma_y$ denotes symplectic matrix with $\sigma_y$ the y-Pauli matrix. A nonzero $E_N > 0$ denotes the existence of bipartite entanglement in our cavity-magnon QED system.

The minimum residual contangle[47, 48] is defined as
$$R_\tau^{min} \equiv \min[R_\tau^{i|jk}, R_\tau^{j|ik}, R_\tau^{k|ij}] \tag{9}$$

where $R_\tau^{b|df} \equiv C_{b|df} - C_{b|d} - C_{b|f} > 0$ $(b,d,f = a_1, a_2, a_3, m_1, m_2, m_3)$, $C_{b|d}$ denotes the squared logarithmic negativity of modes $b$ and $d$. $C_{b|df}$ is the squared logarithmic negativity of modes $b$, $d$ and $f$, which can be calculated by using $\Omega_3 = \oplus_{j=1}^3 i\sigma_y$ instead of $\Omega_2 = \oplus_{j=1}^2 i\sigma_y$, $\tilde{V}_6 = P_{r|st}V_6 P_{r|st}$ ($r,s,t = 1,2,3$) instead of $\tilde{V}_4 = P_{1|2}V_4 P_{1|2}$ in Eq.(8). Here, $P_{1|23} = \text{diag}(1,-1,1,1,1,1)$, $P_{2|13} = \text{diag}(1,1,1,-1,1,1)$ and $P_{3|12} = \text{diag}(1,1,1,1,1,-1)$ are partial transposition matrices. $V_6$ is a $6\times 6$ covariance matrix of three arbitrary modes in our six-mode system, which can be gained by directly deleting rows and columns of irrelevant modes in covariance matrix $V$. Similar to the logarithmic negativity, $R_\tau^{min} > 0$ denotes the existence of tripartite entanglement.

## 3. Results and discussion

First, by solving the set of quantum Langevin equations for semi-classical mean values, we show the mean cavity photon numbers and the mean magnon excitations of our cavity-magnon QED system in Fig. 2.

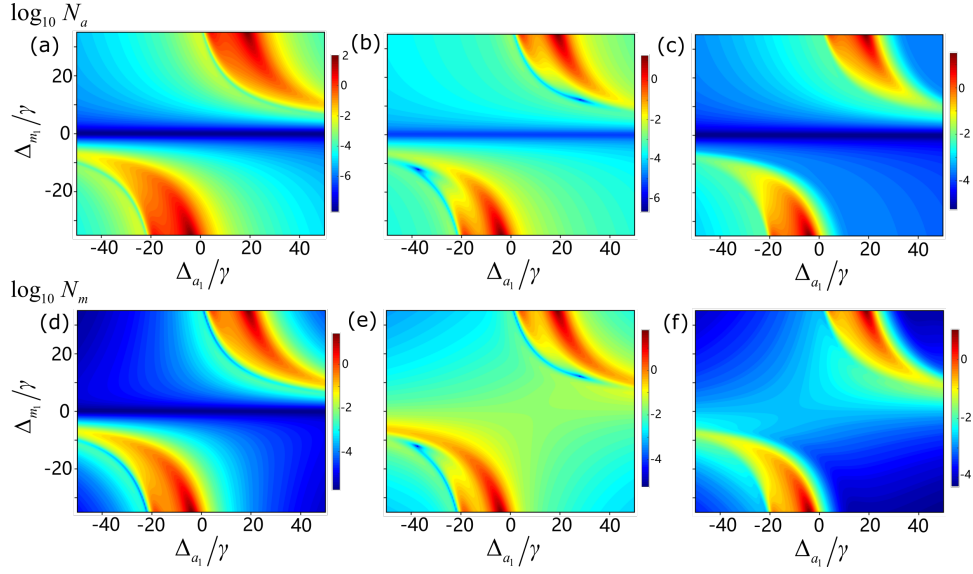

**Fig. 2.** Mean photon numbers of (a) the first cavity $N_{a_1}$, (b) the second cavity $N_{a_2}$, (c) the third cavity $N_{a_3}$, and mean magnon numbers of (d) the first YIG sphere (in cavity 1) $N_{m_1}$, (e) the second YIG sphere (in cavity 2) $N_{m_2}$, (f) the third YIG sphere (in cavity 3) $N_{m_3}$, are plotted on a logarithmic scale as functions of normalized detunings $\Delta_{a_1}/\gamma$ and $\Delta_{m_1}/\gamma$. We set $\Delta_{a_1} = \Delta_{a_2} = -\Delta_{a_3}$, $\Delta_{m_1} = \Delta_{m_2} = -\Delta_{m_3}$, and adopt nonlinear interaction strength $G/2\pi = 4.5\text{MHz}$, linear Rabi frequency $\Omega_2/2\pi = \Omega_3/2\pi = 1\text{MHz}$. See text for the detail parameters.

Mean photon numbers of the first cavity $N_{a_1}$, the second cavity $N_{a_2}$ and the third cavity $N_{a_3}$ are plotted on a logarithmic scale as functions of normalized detunings $\Delta_{a_1}/\gamma$ and $\Delta_{m_1}/\gamma$ in Fig. 2 (a), (b) and (c). Similarly, mean magnon excitation numbers of the first YIG sphere $N_{m_1}$, the second YIG sphere $N_{m_2}$ and the third YIG sphere $N_{m_3}$ are displayed on a logarithmic scale versus normalized detunings $\Delta_{a_1}/\gamma$ and $\Delta_{m_1}/\gamma$ in Fig. 2(d),(e) and (f), respectively. We consider the experimentally realizable parameters[4, 49, 50]: $\omega_{a_j}/2\pi = 10\text{GHz}$, $\omega_{m_j}/2\pi = 10\text{GHz}$, $g_j/2\pi = 20\text{MHz}$, $\kappa_j/2\pi = 5\text{MHz}$, $\gamma_j/2\pi = 1\text{MHz}$ ($j=1,2,3$), $J_{12}/2\pi = 12\text{MHz}$, $J_{23}/2\pi = 12\text{MHz}$, and $T = 20\text{mK}$. We adopt $G/2\pi = 4.5\text{MHz}$, $\Omega_2/2\pi = \Omega_3/2\pi = 1\text{MHz}$, set $\gamma/2\pi = 1\text{MHz}$, and take $\Delta_{a_1} = \Delta_{a_2} = -\Delta_{a_3}$, $\Delta_{m_1} = \Delta_{m_2} = -\Delta_{m_3}$. As shown in Fig. 2, the maximum of mean photon numbers and mean magnon numbers are above 10, but less than 1000. Considering a YIG sphere with diameter 250μm, the total spin number $N = \rho V_m$, where $\rho = 4.22 \times 10^{27} \, m^{-3}$ is the spin density and $V_m$ is the volume of YIG sphere, then the total spin number can be estimated as $N \approx 3.5 \times 10^{16}$. This indicates that the system is in a low-lying excitation state, ensuring the validity of weak excitation assumption $\langle m_j^\dagger m_j \rangle \ll 2N_{m_j} s$. Here, $s = 5/2$ is the spin number of ground-state $Fe^{3+}$ ions of YIG material[11, 40]. Consequently, the Holstein-Primakoff approximation[51] remains applicable.

We then investigate the quantum fluctuations and demonstrate that the cavity-magnon QED system is prepared in an entangled state. Bipartite entanglement quantified by logarithmic negativity $E_{m_1 m_2}$, $E_{m_1 m_3}$, $E_{m_2 m_3}$ and tripartite entanglement quantified by minimum residual contangle $R_\tau^{\min}$ as functions of normalized detunings $\Delta_{a_1}/\gamma$ and $\Delta_{m_1}/\gamma$ were shown in Fig. 3. We adopt nonlinear interaction strength $G/2\pi = 4.5\text{MHz}$, the other parameters are the same as used in Fig. 2. All the results are in the steady state guaranteed by the negative eigenvalues of drift matrix A. As expected, the combined effects of linear cavity-magnon coupling, cavity-cavity coupling and nonlinear interaction from OPA enable the system to be prepared in an entangled state. It denotes that entanglement can be transferred from cavity mode $a_1$ and magnon mode $m_1$ to cavity mode $a_1$ and cavity mode $a_2$, and further to magnon mode $m_2$, cavity mode $a_3$ and magnon mode $m_3$, ultimately resulting in a fully entangled state across all six modes, the indirectly coupled magnon mode and magnon mode get entangled. Notably, the maxima of bipartite and tripartite entanglements occur at different detuning values for photons and magnons.

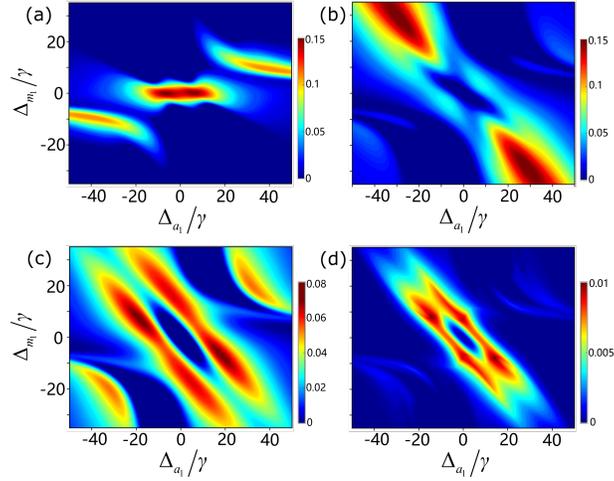

**Fig. 3.** Bipartite entanglement (a) $E_{m_1 m_2}$, (b) $E_{m_1 m_3}$, (c) $E_{m_2 m_3}$ and tripartite entanglement (d) $R_\tau^{\min}$ versus normalized detunings $\Delta_{a_1}/\gamma$ and $\Delta_{m_1}/\gamma$. We take nonlinear interaction strength $G/2\pi = 4.5\text{MHz}$. The other parameters are the same as Fig. 2.

Simultaneously, the variation of bipartite and tripartite entanglements is related to the nonlinear interaction strength ($G$) and the cavity decay rates ($\kappa_1$, $\kappa_2$, $\kappa_3$). We take the detunings $\Delta_{a_1}/2\pi = -10\text{MHz}$ and $\Delta_{m_1}/2\pi = 10\text{MHz}$. Fig. 4 denotes that entanglement exhibits a nonlinear dependence on $G$ and $\kappa_1$. Both bipartite and tripartite entanglements increase with the nonlinear interaction strength $G$. Additionally, Fig. 4 indicates that no entanglement is generated when $G = 0$. This demonstrates that nonlinearity is the primary factor responsible for entanglement generation. Meanwhile, bipartite and tripartite entanglements increase as the cavity decay rates $\kappa_1$ decreasing. The blank regions in Fig. 4 present non-equilibrium states and we find that reducing the nonlinear interaction strength helps maintain the system in a steady state.

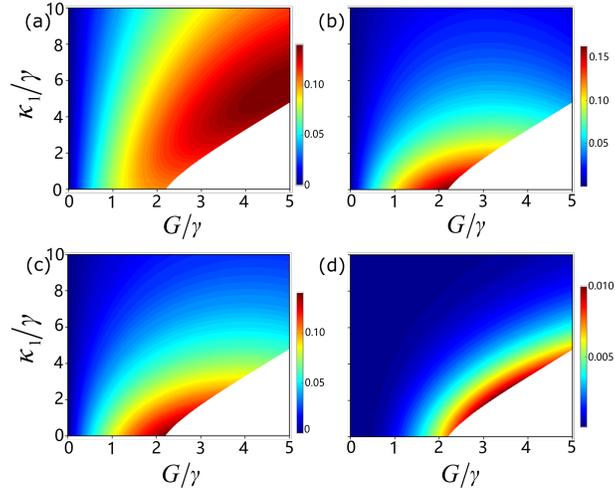

**Fig. 4.** Bipartite entanglement (a) $E_{m_1m_2}$, (b) $E_{m_1m_3}$, (c) $E_{m_2m_3}$ and tripartite entanglement (d) $R_\tau^{\min}$ versus normalized nonlinear interaction strength $G/\gamma$ and decay rate of cavity $\kappa_1/\gamma$. We take $\Delta_{a_1}/2\pi = -10$MHz and $\Delta_{m_1}/2\pi = 10$MHz. The other parameters are the same as Fig. 2.

To enhance both the bipartite and tripartite entanglements, we introduce a second OPA with the same nonlinear interaction strength as cavity 1 and insert it into cavity 2. Bipartite and tripartite entanglements as functions of detunings are shown in Fig. 5. Here, we adopt slight weaker nonlinear interaction strength $G/2\pi = 2.6$MHz to keep the system in steady state. From Fig. 5, we observe that the maximum of bipartite entanglement increases from 0.150 to 0.198 for $E_{m_1m_2}$, from 0.148 to 0.183 for $E_{m_1m_3}$, and from 0.071 to 0.103 for $E_{m_2m_3}$. The maximum of tripartite entanglement $R_\tau^{\min}$ improves from 0.010 to 0.012. It demonstrates that inserting a second OPA is an effective method to enhance entanglement in our QED system.

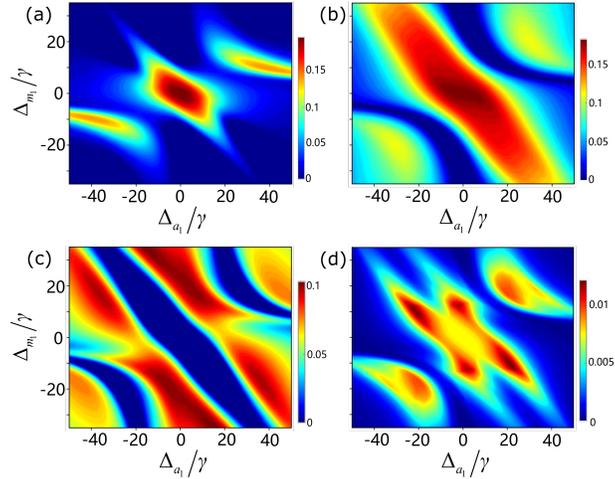

**Fig. 5.** Bipartite entanglement (a) $E_{m_1m_2}$, (b) $E_{m_1m_3}$, (c) $E_{m_2m_3}$ and tripartite entanglement (d) $R_\tau^{\min}$ as functions of normalized detunings $\Delta_{a_1}/\gamma$ and $\Delta_{m_1}/\gamma$ with double OPA inputs. We take nonlinear interaction strength $G/2\pi = 2.6$MHz. The other parameters are the same as Fig. 2.

To further enhance entanglement, we introduce a third OPA inserting to cavity 3 with the same nonlinear interaction strength as cavity 1 and cavity 2. We also adopt slight weaker nonlinear interaction strength $G/2\pi = 2.6$MHz to keep the system in a steady state. Fig. 6 shows bipartite and tripartite entanglements as functions of normalized detunings $\Delta_{a_1}/\gamma$ and $\Delta_{m_1}/\gamma$. We can see that the maximum of bipartite entanglement increases from 0.183 to 0.217 for $E_{m_1 m_3}$, and from 0.103 to 0.201 for $E_{m_2 m_3}$. The maximum of tripartite entanglement $R_\tau^{\min}$ increases from 0.012 to 0.045. It indicates that incorporating a third OPA is an effective method to enhance the overall entanglement in our cavity-magnon QED system. Bipartite entanglement $E_{m_1 m_2}$ decreases from 0.198 to 0.114, the reason may lie in the entanglement partially transferred from $E_{m_1 m_2}$ to $E_{m_1 m_3}$ and $E_{m_2 m_3}$, which denotes the complementary distribution of the entanglement[11]. The setting detunings $\Delta_{a_2}$, $\Delta_{a_3}$, $\Delta_{m_2}$ and $\Delta_{m_3}$ could also contribute to this phenomenon. We will investigate this further in-depth in future research.

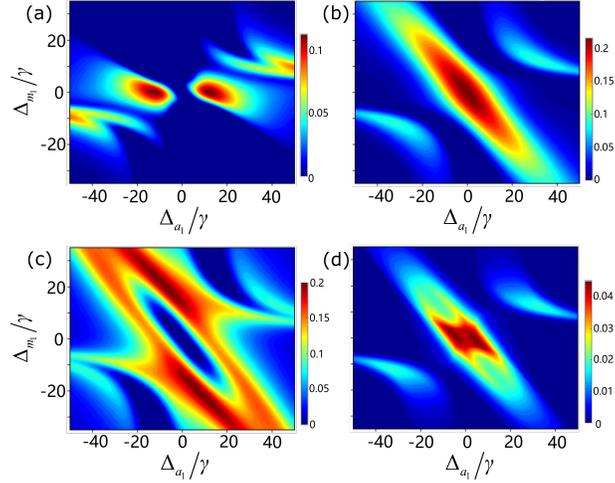

**Fig. 6.** Bipartite entanglement (a) $E_{m_1 m_2}$, (b) $E_{m_1 m_3}$, (c) $E_{m_2 m_3}$ and tripartite entanglement (d) $R_\tau^{\min}$ versus normalized detunings $\Delta_{a_1}/\gamma$ and $\Delta_{m_1}/\gamma$ with triple OPA inputs. We take nonlinear interaction strength $G/2\pi = 2.6$MHz. The other parameters are the same as Fig. 5.

Finally, bipartite entanglement $E_{m_1 m_2}$, $E_{m_1 m_3}$, $E_{m_2 m_3}$ and tripartite entanglement $R_\tau^{\min}$ as functions of bath temperature $T$ are shown in Fig. 7. We take $\Delta_{a_1}/2\pi = 12$MHz and $\Delta_{m_1} = 0$ for $E_{m_1 m_2}$, $\Delta_{a_1} = \Delta_{m_1} = 0$ for $E_{m_1 m_3}$, $\Delta_{a_1}/2\pi = 9$MHz and $\Delta_{m_1}/2\pi = -22$MHz for $E_{m_2 m_3}$, $\Delta_{a_1} = \Delta_{m_1} = 0$ for $R_\tau^{\min}$. It denotes that bipartite and tripartite entanglements are robust against bath temperature and can survive to about 200 mK.

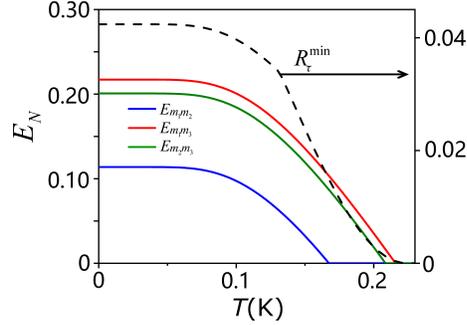

**Fig. 7.** Bipartite entanglement $E_{m_1m_2}$, $E_{m_1m_3}$, $E_{m_2m_3}$ and tripartite entanglement $R_\tau^{\min}$ versus bath temperature $T$. We take $\Delta_{a_1}/2\pi = 12\text{MHz}$ and $\Delta_{m_1} = 0$ for $E_{m_1m_2}$, $\Delta_{a_1} = \Delta_{m_1} = 0$ for $E_{m_1m_3}$, $\Delta_{a_1}/2\pi = 9\text{MHz}$ and $\Delta_{m_1}/2\pi = -22\text{MHz}$ for $E_{m_2m_3}$, $\Delta_{a_1} = \Delta_{m_1} = 0$ for $R_\tau^{\min}$. The other parameters are the same as Fig. 6.

## 4. Conclusion

In conclusion, we have proposed a scheme to prepare bipartite and tripartite entanglements of three magnon modes in a cavity-magnon QED system composed of three cavities, each cavity mode is coupled to a magnon mode in a YIG sphere via linear magnetic dipole interaction, and to other cavity mode via linear coupling interaction. By utilizing the nonlinearity of optical parametric amplifier (OPA), three magnon modes can be entangled with each other, and the steady state of our QED system exhibits genuine tripartite entanglement. We have shown that with experimentally feasible parameters, both bipartite and tripartite entanglements increase with the nonlinear interaction strength of OPA increasing. Furthermore, an alternative entanglement enhancement strategy based on multiplexed OPA inputs was proposed. By implementing individual OPA for each cavity, we observe a significant improvement in entanglement generation. Notably, the achieved entanglement is robust against bath temperature and can survive up to about 200mK. This demonstrates that OPA provides an effective and reliable method for entanglement generation and enhancement in cavity-magnon QED system.

**Acknowledgment.** We thank Jie Li for helpful discussions.
**Disclosures.** The authors declare no conflicts of interest.
**Data availability.** Data underlying the results presented in this paper are not publicly available at this time but may be obtained from the authors upon reasonable request.